\documentstyle[preprint,revtex]{aps} 
\begin{document}
\draft
\begin{title}
The solution to the strong CP problem at the BCS level of chiral rotations.
\end{title}
\author{P. J. de A. Bicudo$^{1,2}$ and J.E. F. T. Ribeiro $^2$}
\begin{instit}
$^1$ Departamento de F\'isica and 
$^2$ Centro de F\'isica das Interac\c c\~oes Fundamentais, 
Edif\'icio Ci\^encia, Instituto Superior Tecnico, Av. Rovisco Pais, 
1096 Lisboa, Portugal
\end {instit}
\begin{abstract}  
We briefly review the cases of forced and spontaneous
chiral symmetry breaking. In particular
the chiral condensate of $q\bar q$ pairs is parametrized with two 
angles, $\phi$ which measures the chiral condensation, and $\theta$ which
measures the chiral rotation. 
The strong CP problem arises when it is assumed that the current
quark masses of the Standard Model, have a $\theta$ phase that 
differs from the phase induced by the instanton
term wich is originated in the QCD sector of the Standard Model.
We show how chiral symmetry 
breaking may solve the strong CP problem at the BCS level. 
We show variationally that the physical vacuum is CP conserving
and therefore  
the interplay of the instanton and the current 
mass terms cannot produce any strong CP violation. 
We also study the possible effect of $\theta$  in weak interactions,
and conclude that it does not produce weak CP.
\end{abstract}
\narrowtext
Within the strong interacting sector of the Standard Model, we have
3 different mechanisms for chiral symmetry breaking. 
\par
First, we have the the lagrangian mass term 
\begin{equation}\label{massa}
{\cal L}_m = m_u \ \bar \Psi_u \ \Psi_u + m_d \ \bar \Psi_d \ \Psi_d 
+ m_s \ \bar \Psi_s \ \Psi_s + \dots
\end{equation}
where the $m_u , m_d , m_s $ current quark masses, explicitely break
chiral symmetry.
\par    
Another well known mechanism for chiral symmetry breaking is furnished
by the 't Hooft ~\cite{Hooft} coupling of the quark flavour determinant to
the instanton,
\begin{equation}
\displaystyle
{\cal L}_g = {C \over g^8}exp({-8\pi^2 \over g^2})
det_{st}\left[\bar\Psi_s(1+\gamma_5)\Psi_t\right] +H.c,
\end{equation}
where $s,t$ are flavour indices. 
\par
Finally the third cause of chiral symmetry breaking is due to the quark
condensate
$\langle \bar \Psi \Psi\rangle $ which is generated dynamically ~\cite{Nambu}.
\par
In what follows we will make use of Valatin-Bogolubov transformations
to construct explicit examples of general orthogonal Fock spaces 
consistent with quark condensation and chiral rotations.
It is this residual freedom to self-consistently choose the Fock space 
that constitutes the cornerstone of the mechanism alowing for the removal 
of CP violating phases.
\par
The condensed vacuum $| \tilde 0 \rangle $ can be generated from the trivial
vacuum $| 0 \rangle $ with a Bogolubov-Valatin transformation, generated by
the parity invariant creator $^3P_0$ of quark antiquark pairs \cite{pap02},
\begin{eqnarray}\label{creator}
&&C^\dagger_{c k}=\sum_{s_1 s_2}b^\dagger_{s_1 c k}
M^\dagger_{s_1 s_2 -\hat k}
d^\dagger_{s_2 c k} \nonumber \\*
&&M^\dagger_{s_1 s_2 \hat k} =  -\sqrt{6} \sum_{m S} \left({ 1 1 0 \atop
 m l 0}\right) \hat k_{1m} \left({ {1 \over 2} \ 
 {1 \over 2} \ 1  \atop s_1 s_2  l}\right) 
 \nonumber \\*
&&M^\dagger= \vec \sigma . \hat k \ \ i \sigma_2
\end{eqnarray}
Then the $| \tilde 0 \rangle $ is given by,
\FL
\begin{eqnarray}
&& | \tilde 0 \rangle =  e^{\displaystyle \sum_{c k}\phi_{k}
\left[C^\dagger_{c k}
-C_{c k}\right]}| 0\rangle   = \prod_{c k}\\
&&       \left[cos^2\phi_{k}+sin\phi_{k}cos\phi_{k} C^\dagger_{c k}+ 
	{sin^2\phi_{k} \over 2} {C^\dagger}^2_{c k}\right]| 0\rangle ,  
	\nonumber
\end{eqnarray}
Now using the Dirac fermion fields,
\begin{equation} 
\Psi={1 \over \sqrt{V}}\sum_{sk} (u_{sk} b_{sk} 
+v_{sk} d^\dagger_{s-k})e^{ik.x}
\end{equation} 
with the chiral spinors normalized with $u^\dagger u=1,\; 
v^\dagger v=1$ in order
to be able to use massless spinors, we get the vacuum expectation value
for the quark condensate, 
\FL
\begin{eqnarray} 
\langle \bar \Psi \ \Psi\rangle =&&
\langle \tilde 0|\bar \Psi \ \Psi |\tilde 0\rangle \nonumber \\
=&& -2\int {d^3 k \over (2 \pi)^3}  
sin\left[-2\phi(k)+atan\left({m/k}\right)\right] 
\end{eqnarray} 
where $[-2\phi(k)]$ is positive in such a way that this expectation value
is always negative and never vanishes. For current quark masses smaller  
than the scale of the strong confining interaction, this
VEV practically has a constant value.
\par
This vacuum is an example of a wider class of vacua which will be denoted 
in this paper by $|\tilde \theta \rangle $.
From the point of view of chiral physics, the extent of chiral symmetry
breaking can be associated to the norm of a vector in a 2 dimensional space.
A cartesian basis of this space can be defined by the "x-direction" of
the scalar condensate $\bar \Psi \Psi$ and by the
orthogonal "y-direction" of the pseudoscalar condensate 
$i \bar \Psi \gamma_5 \Psi $. 
\par  
Therefore it can be said with all generality that the fermion part of the
strong interaction vacuum, must lie somewhere in a circle in this x-y plane
and can be indexed by a polar angle which we denote by $\theta $.
This polar angle parametrizes the chiral rotations generated
by the $Q_5$ operator $\bar \Psi \gamma _0 \gamma_5 \Psi$.
\par
To arrive at the vacuum 
$| \tilde \theta \rangle $ from the trivial unstable vacuum
$| 0 \rangle$ we have to define an extended 
condensate generator in order to
include any combination of scalar 
$^3P_0$ and pseudoscalar $i \ ^1S_0$ 
quark-antiquark pair creations, which we do by replacing
in the equations [\ref{creator}] $M$ by $M_\theta$,
\begin{equation} 
 M^\dagger_\theta = \left[cos(\theta) \vec \sigma . \hat k +i \ sin(\theta)
 \right] i \sigma_2 
\end{equation}
which in turn defines $C^\dagger_\theta$ as,
\begin{equation}\label{criador} 
 C^\dagger_\theta = cos(\theta) C^\dagger_s + sin(\theta) C^\dagger_{ip}.
\end{equation}
In Eq. [\ref{criador}] $C_s$ stands for the $^3P_0$ creator $C^\dagger_{ck}$.  
 $C_{ip}$ represents the pseudoscalar quark-antiquark creator,
where the pseudoscalar spin wave function is $ \uparrow \downarrow
- \downarrow \uparrow = i \ \sigma_2 $. It is also possible 
to rotate directly from the $| \tilde 0 \rangle$ to the 
$| \tilde \theta \rangle$ with the generator $Q_5$. We can show that
we get the algebra,
\begin{eqnarray} 
 [Q_{ip},Q_5]=-2 iQ_{s} \ , \ [Q_{s},Q_5]=2 iQ_{ip}
\end{eqnarray}
\begin{equation} 
 Q_5=\int d^3x\bar \Psi \gamma_5 \Psi =\sum b^\dagger \vec \sigma . \hat k b 
- d^\dagger \vec \sigma . \hat k d \ .
\end{equation}
$Q_5$ generates the chiral rotation of the vacua,
\begin{equation} 
 |\tilde \theta\rangle=e^{i{\theta \over 2}Q_5} |\tilde 0\rangle
\end{equation}
and the effect of this generator is depicted in Fig.[\ref{polar}].
\par
At this stage we just have to bear in mind that all the vacua $|0\rangle$,
$|\tilde 0\rangle$ and $|\tilde \theta\rangle$  are orthogonal and they belong to
independent Fock spaces. For instance,
\begin{equation} 
 \langle\tilde 0|\tilde \theta\rangle=\prod_{c k}
 \left[1-sin^2({\theta \over 2})sin^2(2 \phi) \right]    \
 =\delta_{0,\theta}
\end{equation}
Thus all these vacua are independent and cannot be reached pertubatively.
Each vacuum has its own field creators 
$b^\dagger_{\displaystyle\theta_{sck}}, 
d^\dagger_{\displaystyle\theta_{sck}}$ 
and field annihilators $b_{\displaystyle\theta_{sck}}, 
d_{\displaystyle\theta_{sck}}$ 
that can be used to generate the corresponding Fock spaces ${\cal F}_\theta$.
\par
When there is only one source of chiral symmetry breaking,
like ${\cal L}_m$, then we can trivially choose this vacuum to define the
x-axis $(\theta =0)\;:|\tilde 0 \rangle $
\par 
When a second source for chiral symmetry breaking exists, like for
instance ${\cal L}_g$, then we have an extra direction and a P violating
phase could be thouht to appear. To quote 't Hooft
~\cite{Hooft},
 "if other mass terms or interaction terms occur in the
Lagrangian that also violate chiral U(1), then they may have a phase factor
different from these. We then find that our effective Lagrangian may violate
P, whereas C invariance is maintained."
\par
This is the origin of the strong CP problem, in the sense that 
strong CP violations have not yet been observed experimentally. In
other words, because weak and strong interactions are independent, and in a 
sense one is skewed in relation to the other~\cite{skew}, the angles 
$\theta_{{\cal L}_m}$ ( which we had set to be the origin)
and $\theta_{{\cal L}_g}$ are naturally expected to be different. An arbitrary 
example of this effect is shown in Fig. [~\ref{polar}].
\par
The purpose of this paper is to show that a difference in the angles
$\theta_m$ and $\theta_g$ may not cause strong CP violation.
\par
If there were no explicit chiral symmetry breaking Lagrangians
${\cal L}_m$ and ${\cal L}_g$ then a continuous
degenerate set of orthogonal vacua would exist, and the whole hadronic
physics could suffer a chiral rotation from the vacuum $| \tilde 0 \rangle$ to
another vacuum $| \tilde \theta \rangle$. For each $\theta$ we would find
an identical replica of the whole hadronic world. Nothing in strong 
interaction physics will distinguish one vacuum from another.

This is easy to see in a simplified version of ${\cal L}_g$ 
provided by the Lagrangian of Baluni~\cite{Baluni} with only 1 flavor.
Let us suppose that we have the following Lagrangian,
\begin{eqnarray} \displaystyle
{\cal L}_m + {\cal L}_g +{\cal L}_i= m\ &&\bar \Psi \ \Psi  
+ \bar \Psi (A + i B \gamma_5) \Psi +{\cal L}_i\nonumber \\
= m^* \ \bar \Psi &&e^{\displaystyle (i \theta \gamma_5)} \Psi 
+{\cal L}_i\nonumber \\
m^* = \sqrt{(m+A)^2+B^2} &&\ , \ \
\theta= arctan{B \over m+A}
\end{eqnarray}
where ${\cal L}_i $ contains the remainder of the Q.C.D. Lagrangian 
which is chiral invariant and should not concern us for the purpose 
of this paper.  We can use the Valatin-Bogolubov 
transformation,
\begin{eqnarray}
\Psi_\theta=e^{i Q_5 \frac{\theta}{2}}\; \Psi e^{-i Q_5 \frac{\theta}{2}}=
e^{i \gamma_5 \frac{\theta}{2}}\; \Psi \ , \nonumber \\
\Psi_\theta = {1\over \sqrt{V}} \sum u \ \ b_\theta + v \ \ d^\dagger_\theta
\end{eqnarray}
to get read of the complex CP violating masses. Thus if instead of the 
fields $\Psi$, we use the chiral 
rotated fields $exp(i{\theta \over 2} \gamma_5)\Psi$ then CP violation 
explicitely disappears from the Lagrangian for the vacuum  
$|\tilde \theta\rangle=e^{i Q_5 \frac{\theta }{2}}\; |\tilde 0\rangle $. 
It is not difficult to show
that this choice minimizes the vacuum energy, 
\begin{equation}
{\cal E}= \langle \bar \Psi \Psi \rangle \left[(m+A)\,cos(\theta)  
+B\,sin(\theta)\right]+{\cal E}_i \ .
\end{equation}
$|\tilde \theta \rangle$ is therefore the physical vacuum of
the system, and for this physical vacuum there remains no evidence of 
strong CP. 
\par
For the purpose of
this letter it suffices to calculate the vacuum expectation value
of the hamiltonian density, which can be performed with the help
of the Wick theorem. 
\par
Let us define a class of 't Hooft determinants for n flavours as,
\begin{equation}\label{hooft}
D_n={1 \over 2} \left[ det_{st}\bar s(1+\gamma_5)t 
+det_{st}\bar s(1-\gamma_5)t \right]
\end{equation}
where we drop the $\Psi$ and from now on $s$ , $t$ represent 
the Dirac fields of flavour $s$, $t$.
In Eq. (\ref{hooft}) only the even terms in $\gamma_5$ wich are CP conserving 
survive.
In fact this is simply equivalent to insert an even number of $\gamma_5$ 
matrices in the flavour determinant $det_{st}\bar s t$ in 
all possible different ways. 
For instance we get the mass term for 1 flavour, whereas for 2 flavours
we already get 4 terms, while for 3 flavors we would have 24 terms and so on,
\begin{eqnarray}
D_1 =&& \bar u u     \nonumber \\
D_2 =&& 
\bar u u \  \bar d d  
- \bar u d \ \bar d u 
+ \bar u \gamma_5 u \ \bar d \gamma_5 d 
- \bar u \gamma_5 d \ \bar d \gamma_5 u 
 \nonumber \\
D_3 =&& \bar u u \  \bar d d \  \bar s s + \dots    
\end{eqnarray}
Let us see how $D_n$ transforms with a
general chiral rotation of angles 
$\theta_u \ , \ \theta_d \ , \ \dots$ . We find, 
\FL
\begin{eqnarray}
D_n\rightarrow&& D_n^{\theta_u,\theta_d, \dots} \nonumber \\
=&&{1 \over 2}  e^{i(\theta_u+\theta_d + \dots)}
det_{st}\bar s(1+\gamma_5)t   \nonumber \\ 
&&+{1 \over 2} e^{-i(\theta_u+\theta_d + \dots)}
det_{st}\bar s(1-\gamma_5)t   \\ 
=&&cos(\theta_u+\theta_d + \dots)D_n+ 
i \ sin(\theta_u+\theta_d + \dots)D^5_n    \nonumber
\end{eqnarray}
Where $D^5_n$ has the same definition as $D_n$ except
that it has an odd number of insertions of $\gamma_5$.
While $D_n$  is CP invariant, $D^5_n$ is a pseudoscalar. Because
all the angles come in a sum, we see that from the point of
view of chiral rotations $D_n$ has a single dimension, and 
thus is equivalent to a mass term in the flavour U(1) direction.
\par
In order to study the vacuum energy we have to calculate 
the vacuum expectation values of the operators $D_n$. This can be 
performed with the Wick contraction technique. 
As an illustration we give $\langle D_n \rangle$ for the simpler 
case where all the flavours have the same condensate. This is not
a bad approximation for the u, d quarks and of course this can also
be calculated in the general case of unequal quark condensates.
\FL
\begin{equation}
\langle D_n\rangle=
\langle\tilde 0|D_n |\tilde 0\rangle= G \langle\bar \Psi \ 
\Psi\rangle^n  
\end{equation}
where G is a geometrical factor, 
\begin{equation}
G_1=1 \ , \ G_2={3 \over 2} \ , \ G_3= 3 \ , \ \dots
\end{equation}
\par
In the general case we get for the vacuum energy density 
\begin{eqnarray}\label{energy}
&&\langle\tilde \theta_u \tilde \theta_d \dots |D_n |
\tilde \theta_u \tilde \theta_d \dots \rangle = 
cos(\theta_u+\theta_d+\dots) \langle D_n \rangle  \nonumber \\
&&{\cal E}={\cal E}_i+m_u\langle\bar u u\rangle 
cos(\theta_u) + m_d\langle\bar d d\rangle 
cos(\theta_d)  
\nonumber \\
&& +\dots+ K \langle D_n \rangle 
cos(\theta_u+\theta_d+\dots-\theta_g) 
\end{eqnarray}
where $\langle D_n \rangle$ is a given number and $K$ is given by
\begin{equation}
K=2 {C \over g^8}exp({-8\pi^2 \over g^2}) 
\end{equation}
\par
In the litterature the $\theta_g$ is regarded as fixed. However for 
completness we will also consider the scenario where $\theta_g$ is 
a variational parameter. This is an $easy$ scenario insofar that the
evident solution is that $\theta_g$ will be "aligned"
with the current mass chiral angles, $\theta_f=\theta_g=0$. This 
is the combination
that yields the lowest vacuum energy. In this case the instanton Lagrangian
does not produce any CP violation either in strong or in weak 
interactions. 
\par
Now we will suppose that $\theta_g=\theta$ is fixed and show that even 
in this case  strong CP violation does not occur.
The minimum condition is simply obtained when we differentiate 
Eq.[\ref{energy}] with respect to the $\theta_f$, and we get the set
of coupled equations,
\FL
\begin{eqnarray} \label{minimum}
&m_u\langle\bar u u\rangle sin(\theta_u)=-sin(\theta_u+\theta_d+\dots-\theta)K\langle D_n\rangle & 
\nonumber \\
&m_d\langle\bar d d\rangle sin(\theta_d)=-sin(\theta_u+\theta_d+\dots-\theta)K\langle D_n\rangle& 
\nonumber \\
&\dots& 
\end{eqnarray}
that can be easilly solved numerically.
For small chiral angles and small current masses
the system becomes linear an we get the simple
solutions,
\begin{equation}
\theta_f= \theta \ m_f^{-1} / \sum_{s} m_s^{-1} \ .
\end{equation}
Thus only the lightest flavours are actually rotated because this
corresponds to an average weighted by the inverse masses. If we inspect
the light $ \pi, K, \eta, \eta'$
pseudoscalar meson masses with the help of the Gell-Mann, Oakes
and Renner relations, together with the heavy meson masses, we get the
approximate mass ratios, 
\FL
\begin{eqnarray}
& m_d \simeq 2 m_u \ , \ m_s \simeq 30 m_d  < \ K\langle D_n\rangle/\langle\bar \Psi \Psi\rangle
\nonumber \\
& << \ m_c \simeq 6 m_s \ , \ m_b \simeq 3 m_c  \ , \ m_t \simeq 15 m_b
\end{eqnarray}
which in turn yield the chiral rotations,
\FL
\begin{equation}
\theta_u \simeq {2 \over 3}\theta \ , \theta_d \simeq {1 \over 3 }\theta 
\ , \theta_s \simeq {1 \over 90 }\theta \ , \
\theta_c \simeq \theta_b \simeq \theta_t \simeq 0 
\end{equation}
\par
We now verify that within the minimum energy vacuum $|\tilde \theta\rangle$ 
the Lagrangian becomes CP invariant. We have,
\begin{eqnarray}
D_n= && \langle D_n\rangle  +\bar u u\langle D_n^u\rangle  +\bar d d\langle D_n^d\rangle +\dots \nonumber \\
&& + higher \ order \ terms
\end{eqnarray}
Where $\langle D_n^f\rangle={\langle D_n\rangle \over \langle\bar f f\rangle}$ is simply equal to 
$\langle D_n\rangle$ except for a contraction of 
$\bar f f$ or $f \bar f$. In the same way 
we get,
\begin{eqnarray}
D_n^5= && 0 + \bar u \gamma_5 u\langle D_n^u\rangle  +\bar d \gamma_5 d\langle D_n^d\rangle +\dots 
\nonumber \\ && + higher \ order \ terms
\end{eqnarray}
thus we verify, with the help of equation [\ref{minimum}] 
that in the normal ordered hamiltonian all the CP
violating terms cancel up to quadratic terms in field operators.
\FL
\begin{eqnarray}
&&{\cal H}={\cal E} + {\cal H}_{{\displaystyle i}2} + \nonumber \\*
 &&+ \left[ m_u cos(\theta_u) + cos(\theta_u + \theta_d + \dots -\theta) K 
 \langle D_n^u\rangle \right] \bar u \ u  \nonumber \\*
 &&+ \left[ m_d cos(\theta_d) 
 + cos(\theta_u + \theta_d + \dots -\theta) K  \langle D_n^d\rangle \right] \bar d \ d    
 \nonumber \\&&+ \dots  \nonumber \\*
 &&+ i\left[ m_u sin(\theta_u) + sin(\theta_u + \theta_d + \dots -\theta) K 
 \langle D_n^u\rangle \right] \bar u \gamma_5 u  \nonumber \\*
 &&+ i\left[ m_d sin(\theta_d) 
 + sin(\theta_u + \theta_d + \dots -\theta) K  \langle D_n^d\rangle \right] 
 \bar d \gamma_5 d    \nonumber \\*
 &&+ \dots  \nonumber \\  && + higher \ order \ terms \nonumber \\*
&&={\cal E} + {\cal H}_{{\displaystyle i}2}
+ m^*_u\bar u \ u 
+ m^*_d\bar d \ d + \cdots \nonumber \\
&&
+ i\;[0]\;\bar u \gamma_5 u 
+ i\;[0]\;\bar d \gamma_5 d + \cdots + higher \ order \ terms 
\end{eqnarray}
A word of caution is in order.
It is well known that the BCS formalism does not exactly minimize the vacuum
energy. To solve this problem  
one has to include the effects of coupled channels in the vacuum 
and this will be not pursued here. 
However the BCS approach is already enough to show the cancellation
of the CP violating quadratic terms considered in the litterature
\cite{Baluni}.
\par
While strong CP violation disappears, 
a chiral rotation of the quark fields might
affect weak interactions. The Cabibo-Kobayashi-Maskawa ~\cite{Maskawa}
matrix, responsible for 
the weak decay of quarks is not invariant for chiral transformations
that are flavour dependent. In the usual Standard 
Model formalism, it appears in the form,
\begin{equation}
{\cal L}_{C. K. M.} = {g \over 2}\left(  
W_\mu^+\sum_{s,f} \bar {u_{L s}} \gamma^\mu V_{s f} d_{L f} + H.c.
\right)
\end{equation}
where $u_s$ corresponds to the Dirac field 
for the flavours $u, c, t$ and $d_f$ corresponds to the the Dirac fields
for the flavours $d, s, b$.
In the CKM matrix a flavour dependent chiral rotation would produce 
the phase  $(\theta_s - \theta_f)/2$ 
for the terms  with $\bar u_{Ls} \gamma^\mu d_{Lf}$ and the opposite
phase for the hermitean conjugate terms $\bar d_{rf} \gamma^\mu u_{rs}$. 
This $\theta$ phase is independent of the single 
CP violating phase  of the standard model.
For instance in the case where only the light flavours suffer a chiral
rotation 
to their strong CP invariant values $\theta_f$,
$V$ will be transformed into
\FL
\begin{eqnarray} \displaystyle
V \rightarrow V'=\left(\begin{array}{ccc}
V_{ud} e^{i(\theta_u-\theta_d)\over2}\ & \ V_{us} e^{i(\theta_u-\theta_s)\over2} \ & 
\ V_{ub} e^{i\theta_u\over2}\\
V_{cd} e^{-i\theta_d\over2} \ & \ V_{cs} e^{-i\theta_s\over2} \ & 
\ V_{cb} \\
V_{td} e^{-i\theta_d\over2} \ & \ V_{ts} e^{-i\theta_s\over2}\ & \ 
V_{tb} 
\end{array}\right).
\end{eqnarray}
It is easy to verify that if $V$ was originally unitary then $V'$ is also 
unitary. The new phases $\theta_f/2$ cancel, and we get,
\begin{equation}
V'^\dagger V' = V'^\dagger V' = 1
\end{equation}
Moreover the new phases can be rotated out of the CKM matrix if we rotate
each flavour $f$ with a simple phase $e^{-i \theta_f}/2$, which is not a chiral 
phase and will
vanish in any other terms in the Lagrangian where Dirac fields always come in 
pairs $\bar f \dots f$.
To verify that this mechanism is not a
candidate to explain weak CP, then we see that the products
\begin{equation}
Im[V'_{ud}   \ V'^*_{us} \ V'^*_{cd} \ V'_{cs} ] =
Im[V_{ud}   \ V_{us}^* \ V_{cd}^* \ V_{cs} ]
\end{equation}
that mesure weak CP violation are invariant with regard to the phases
$\theta_f$. We thus conclude that the mechanism that  preserves strong
CP invariance does not affect the CP of weak interactions.
\par
An interesting conclusion of this work is that the $\pi_0$ valley can be
seen in the direction $\theta_d=\theta -\theta_u$  between 
the $\eta$ twin peaks as we show in Fig.[\ref{coseno}]. The $\eta$ direction
is defined with $\theta_d=\theta_u$, and the curvature is essentially
forced by the instanton term. We thus can extend the Gell-Mann Oakes
Renner relation to the $\eta$ meson providing we replace $m_u\langle\bar u u\rangle$
by $K\langle D_n\rangle$ in order to get, in the limit of small $\theta$,
\begin{equation}
K\langle D_n\rangle \simeq m_{\eta}^2 f_{\pi}^2
\end{equation}
\par 
We would like to thank our colleages in CFIF for interesting discussions, 
especially Prof. Gustavo Branco, Dr. Walter Grimus and Dr. Luis Lavoura. 

\figure{
We show the chiral of ${\cal L}_m$, ${\cal L}_g$. For historical
reasons we choose a vanishing $\theta_m$. 
The physical $\tilde \theta$ is also visible.
\label{polar}}
\figure{
Here we show how the vacuum energy (in arbitrary units) 
changes when 2 different flavours 
are rotated. 
In this example we have $\theta=\pi / 3$. 
The $\pi$ valley is in the middle of 
the $\eta$ hills, with a visible saddle point.
\label{coseno}} 
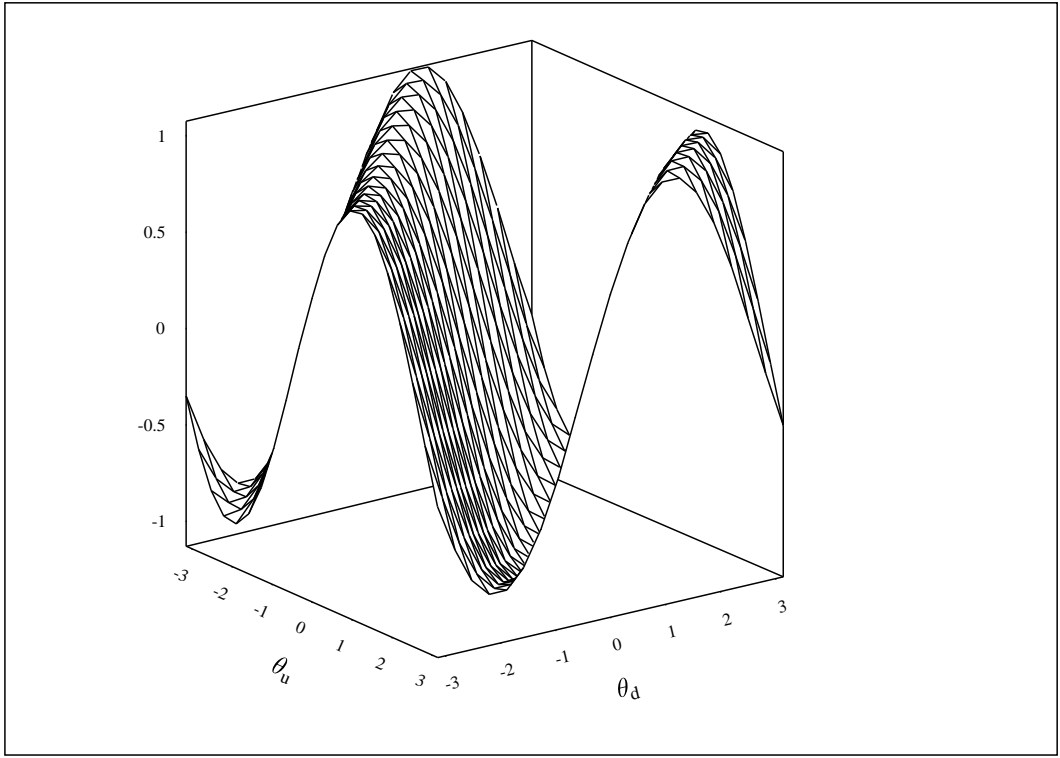
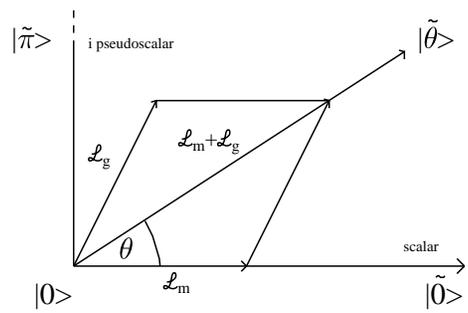
\end{document}